\documentclass{aastex631}
\submitjournal{AJ}

\shorttitle{AASTeX v6.3.1 Sample article}
\shortauthors{Zhao C. et al.}
\graphicspath{{./}{figures/}}

\begin{document}
\title{An Automatic Approach for Grouping Sunspots and Calculating Relative Sunspot Number on SDO/HMI Continuum Images}

\correspondingauthor{Shangbin Yang}
\email{yangshb@bao.ac.cn}

\author[0009-0003-3573-4514]{Cui Zhao}
\affiliation{College of Applied Science and Technology, Beijing Union University, Beijing 102200, PR China}

\author[0000-0002-2967-4522]{Shangbin Yang}
\affiliation{National Astronomical Observatories, Chinese Academy of Sciences, 20A Datun Road, Chaoyang District, Beijing 100012, PR China}

\author{Tingmei Wang}
\affiliation{College of Applied Science and Technology, Beijing Union University, Beijing 102200, PR China}

\author{Haiyan Zhao}
\affiliation{College of Applied Science and Technology, Beijing Union University, Beijing 102200, PR China}

\author{Shiyuan Liu}
\affiliation{College of Applied Science and Technology, Beijing Union University, Beijing 102200, PR China}

\author{Fangyuan He}
\affiliation{College of Applied Science and Technology, Beijing Union University, Beijing 102200, PR China}

\author{Zhengkun Hu}
\affiliation{College of Applied Science and Technology, Beijing Union University, Beijing 102200, PR China}



\begin{abstract}

Relative Sunspot Number is one of the major parameters for the study of long-term solar activity. The automatic calculation of the Relative Sunspot Number is more stable and accurate as compared to manual methods. In this paper, we propose an algorithm that can detect sunspots and divide them into groups, to automatically calculate the Relative Sunspot Number. Mathematical Morphology was adopted to detect sunspots then group them. The dataset used were the continuum images from SDO/HMI. The process was carried out on the overall HMI data available on the timespan from January 2022 to May 2023 with a time cadence of one day. The experimental results indicated that the method achieved high accuracy of 85.3\%. It was well fitted with the international Relative Sunspot Number provided by Solar Influences Data Analysis Center (SIDC) (CC=0.91). We calculated the conversion factor K value of SDO/HMI for calculating the Relative Sunspots Number(K=1.03).

\end{abstract}

\keywords{Solar, Relative Sunspot Number, sunspots groups, Astronomy image processing}


\section{Introduction} \label{sec:intro}
The Relative Sunspot Number(SN) represents the strength of solar activity and provides the longest record of solar activity\citep{clette2014revisiting, clette2015revision}. Coronal mass ejections and strong explosions on the solar surface can be observed more frequently when the sunspot number reaches its maximum. Continuous observation of sunspots has made it possible to monitor solar activity and thus predict the space environment\citep{hathaway2004sunspot, lukianova2011changed, yan2011phase}. SN has long also been used in studies of solar dynamo, predictions of Solar cycle, and studies of secular variation of the Earth's climate\citep{solanki2004unusual, wang2004sun}.

The SN was first recorded according to the equation proposed by Wolf at the Zurich Observatory in 1849, and later taken over by the Solar Influences Data Analysis Center (SIDC)\citep{vanlommel2004sidc,clette2007wolf}. The SN is defined by multiplying the number of observed sunspot groups by 10, adding the total number of observed sunspots to the product, and then multiplying by the parameter K\citep{hossfield2002history, clette2016revised}. K is a conversion factor, which stipulates that the K of Zurich Observatory is 1. The values of K for the other observatories are determined by comparing their observations with the Zurich Observatory's Sunspot Number. The K value is related to factors such as weather conditions (transparency and Astronomical seeing) at the observatory, the aperture of the telescope and the experience of the observer. Different observation stations or instruments have different K values.

Conventionally, the calculation of SN is manual based and due to human subjectivity, the results are not stable. In recent years, with the development of image processing technology and deep learning technology, a variety of automatic sunspots extraction technologies have been proposed \citep{curto2008automatic, zhao2016automatic, carvalho2020comparison, hanaoka2022automated}. The statistical accuracy of the total number of sunspots has gradually improved. However, there are few automatic methods for sunspots grouping, the accuracy is also very low. For instance, \cite{dasgupta2011sunspot} used clustering algorithm to group sunspots, and compared their results with Wolf Number. The difference error between the two was about 10\%. \cite{palladino2022sunspot} used the deep learning technology to group sunspots, and the accuracy of sunspot group recognition was 44.22\%.

As the new solar observation instruments have collected massive sunspot data, a stable and accurate approach to accurate the relative SN is needed for these data. In this paper, we proposed an algorithm that could detect and group sunspots so that the Relative Sunspot Number could be  calculated automatically. The dataset used in the study were SDO/HMI Continuum images. We further compared the calculated SN with manual means, Catalogue of Heliophysics Features (HFC) and SIDC respectively. In addition, the K value of the HMI was also calculated. To our knowledge, it is the first time that the K value for the HMI has been calibrated. The paper is organized as follows: Section 2 focuses on data. The proposed model is given in Section 3. Section 4 presents the results and discussion. Section 5 provides the method to calculate the K value. Finally, the summary is provided in Section 6.

\section{Data} \label{sec:data}
We used the continuum images from SDO/HMI as the dataset. SDO/HMI is a space device, conducts continuous observation of the day 24 hours without interruption. It started the regular observation on April 30, 2010, providing high-quality sunspot full-disk data. It publishes a set of data every 15 minutes, including magnetic field data and corresponding continuous images. There are four different scale size of images, including: 256, 512, 1024, and 4096. We selected the daily images at 08:45:00UT from January 2022 to May 2023, with the image size of 1024 and an interval of one day. There were 507 images included in total. The original images were RGB, and converted to Gray by the standard Matlab program rgb2gray().

\section{Methods} \label{sec:methods}
As SN is calculated by the number of sunspots and sunspot groups, the method proposed in this paper consists of two pipelines: the first part is an automatic sunspot recognition method for automatically detecting sunspots and calculating the total number of sunspots on the solar disk. The second part is an automatic grouping method of sunspots, for automatically calculating the number of sunspot groups.

Despite the rapid development of deep learning and computer vision technology used in image processing, here we still chose the traditional image morphology processing techniques. It is mainly based on the following considerations: (1) For images with uniform background, the target and background grayscale are significantly different, the traditional Mathematical Morphology(MM) generally has good performance. As sunspots are obviously different from the solar disk, MM should be suitable for detecting sunspots. (2) Deep learning technology requires a large amount data annotation and high computational performance, which requires a lot of manpower and computational power. While traditional image processing technologies do not impose these requirements.

\subsection{Mathematical Morphology Algorithm} \label{subsec:morphology}
Mathematical morphology is a discipline of image analysis based on lattice theory and topology, which studies spatial structure and morphology \citep{serra1982image, heijmans1995mathematical}. The basic concept of it is to change the shape and features of an image by performing specific operations with the structural elements on the images. The commonly used structuring elements are crosses, squares, and open disks.

The basic morphological operators are erosion and dilation. If we set a binary image as $\emph{J}:u \rightarrow {0,1}$, and the foreground pixels are $u_f = J^{-1} (\{1\})$. The erosion and dilation of the binary image \emph{J} by the structuring element $S \in Z \times Z$ are defined as:
\begin{equation}
    Erosion:\mu_f \ominus S= \{x|\forall_s \in S, x+s \in \mu_f\}
\end{equation}
\begin{equation}
   Dilation: \mu_f \oplus S= \{x + s|x \in \emph{J} \land s \in S\}
\end{equation}

The other two major operations in morphology are opening and closing. Opening operation is considered to be erosion, followed by dilation and this operation eliminates small objects and sharpens peaks in the object. Closing operation is first dilation and then erosion, which fuses narrow breaks and fills small holes and gaps in the image.

Dilation is an operation that looks for local maximum and makes the highlighted area larger. Erosion is the operation that looks for the local minimum and makes the highlighted area smaller. For sunspot images, the highlighted part is the solar disk, and the black parts are the sunspots. As the dilation and erosion operations are two dual operations, the dilation and erosion operations can be applied to the black area as well. Dilation makes sunspots smaller, and erosion makes sunspots larger.

\subsection{Automatic recognition of sunspots} \label{subsec:recognition}
We described the automatic sunspot identification method in detail in \cite{zhao2016automatic}, and here we briefly describe the processing flow with the SDO/HMI continuum image of January 14, 2022 as an example:

(1) For the observation image shown in Figure. 1(a), the morphological closing operation was applied to eliminate sunspots and other darker areas on the solar surface, obtain a clean solar disk, as shown in figure 1(b). The shape of the structuring element used in closing is disk, with a size of 10.

(2) Figure 1(a) was subtracted from Figure 1(b) yield the darker regions on the solar disk, as shown in Figure 1(c).

(3) Figure 1(c) is segmented automatically by the threshold to obtain the Figure 1(d), which was a binary image with white areas as candidate sunspots. Considering the limb darkening, the threshold for the area within [1:1000,100:900] was set to 22, and the rest was set to 15.

(4) To eliminate the noises of the solar disk that can be easily regarded as sunspots (which may be instrument-generated noise or darker areas on the solar disk), we set the area of each candidate region to be greater than a fixed value of $5$. In addition, since the gray value of the noise is relatively uniform,  and the gray value of the sunspots center are lower than that at the edge, we set that the difference between the maximum and minimum values of pixels in the set candidate area needs to be larger than a fixed value of $5$. With the above rules, the noises on the solar surface were removed and the pure sunspot regions were obtained, as presented in Figure 1(e). The sunspots are marked in red.

\begin{figure*}
\gridline{\fig{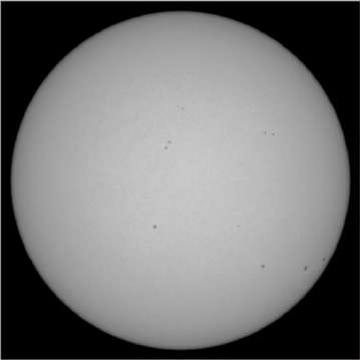}{0.35\textwidth}{(a)}
          \fig{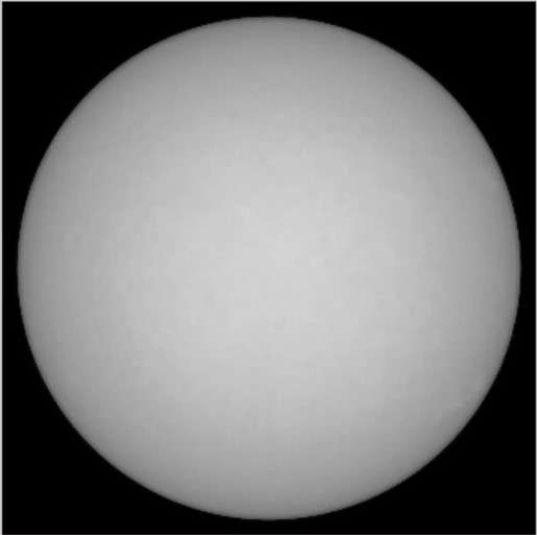}{0.35\textwidth}{(b)}
          }
\gridline{\fig{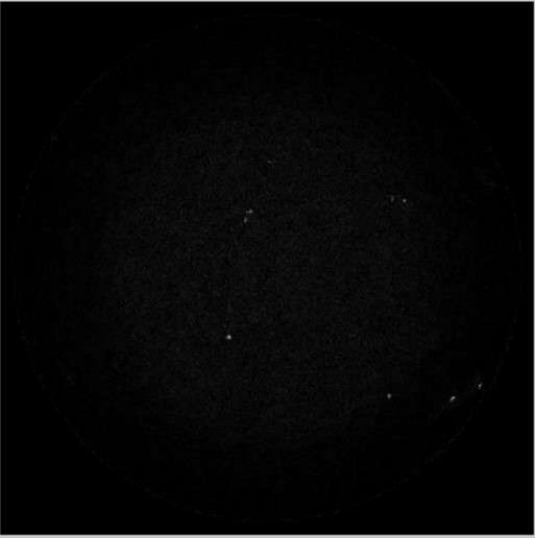}{0.35\textwidth}{(c)}
          \fig{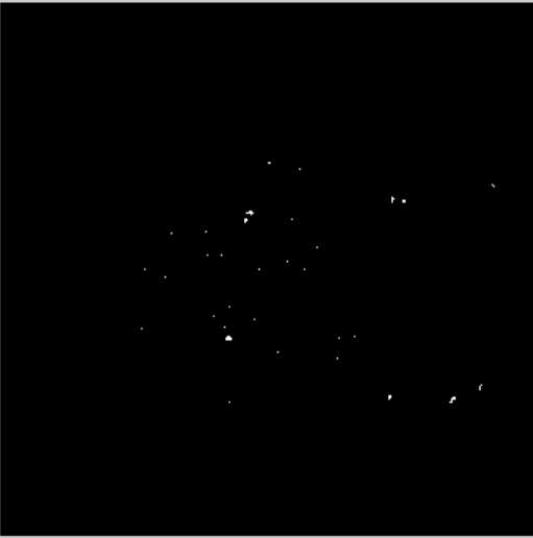}{0.35\textwidth}{(d)}
          }
\gridline{\fig{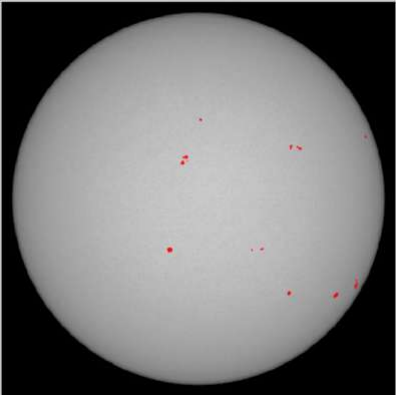}{0.35\textwidth}{(e)}}
\caption{An example of continuum image on January 14, 2022 to carry out automatic sunspots recognition: (a) the original image, some sunspots are seen on the solar disk. (b) the result of performing a closing operation on (a). (c) the result of subtracting (a) from (b). (d) a binary image, which is the result of the threshold automatic segmentation of (c). (e) sunspots marked with red color are detected on the disk.
\label{fig:figone}}
\end{figure*}

\subsection{Group sunspots and calculate the relative SN automatically} \label{subsec:grouping}
Since sunspots belonging to a group are generally clustered together, we assume that the sunspots smaller than a certain distance should be classified as a group. Therefore we designed the algorithm flow shown in Figure 2:

(1) The white regions in Figure 2(a) are the sunspots detected in the first stage. Morphological Erosion operation was carried out on Figure 2(a), and a suitable structural element of size $15$ was selected experimentally. The white areas closer to each other merged into a large area, as shown in Figure 2(b).

(2) Each white region in Figure 2(b) was labelled and numbered with a red rectangle, and then superimposed on the original Figure 1(a) to obtain the sunspots grouping Figure 2(c). This step was made automatically by Matlab programs. It was  divided into three steps: Firstly, the program regionprops() was used to get each white area bounding box. It was the smallest rectangle containing the white area, which was extracted by the program automatically. Secondly, the program rectangle() was applied to show it out. Meanwhile the program text() was used to label it. At last, the bounding boxes were superimposed to the original image.

(3) To show the grouping effect clearly, we enlarged the sunspot groups numbered 3 and 9, as shown in Figure 2(d) and 2(e). It could be seen that the smaller and darker sunspot groups and sunspot groups at the edge of the solar disk have been identified, and the algorithm was more effective in identifying sunspot groups.

As shown in the figure below, the total number of sunspots was 15 and the number of sunspot groups was 9, so the relative number of sunspots was 105.

\begin{figure*}
\gridline{\fig{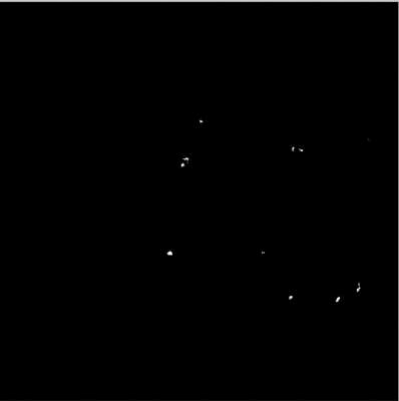}{0.3\textwidth}{(a)}
          \fig{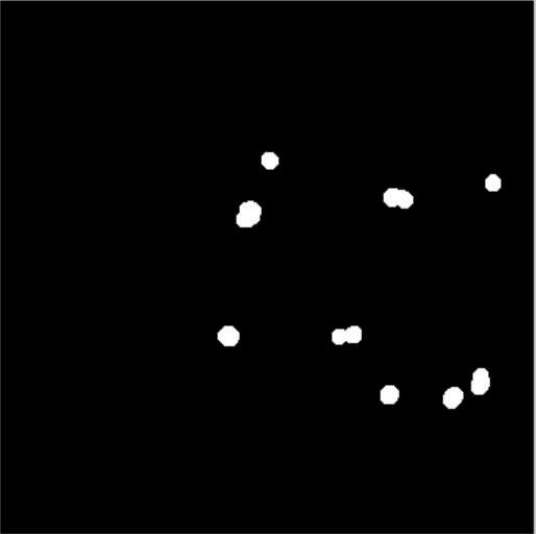}{0.3\textwidth}{(b)}
          }
\gridline{\fig{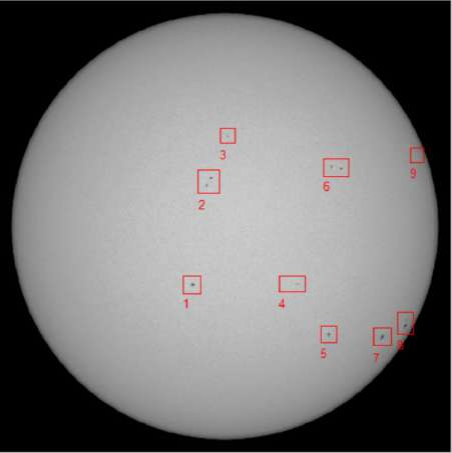}{0.4\textwidth}{(c)}}

\gridline{\fig{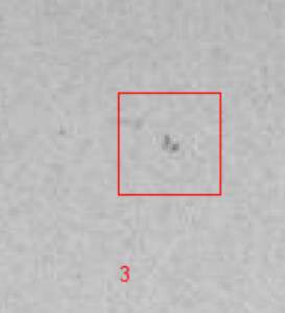}{0.2\textwidth}{(d)}
          \fig{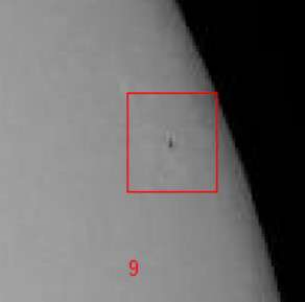}{0.2\textwidth}{(e)}
          }

\caption{Group sunspots after they are be detected: (a) the sunspots detected in the first stage. (b) the results of erosion calculation on figure (a). (c) sunspots grouping labeled with a total 9 groups and overlayed to the original image. (d) enlarge and display group 3. (e) enlarge the display group 4.
\label{fig:figtwo}}
\end{figure*}

\section{Results} \label{sec:results}
\subsection{Compare our results with HFC's} \label{subsec:results1}
Catalog of Heliophytics Features (HFC)(\cite{bonnin2013heliophysics}) is the Solar Survey Archive of BASS2000 available at \url{https://bass2000.obspm.fr/home.php}. It provides daily sunspot groups numbers, which were adopted here as a comparison with our method.

Since the HFC can only select a particular date to search for the number of sunspot groups on that day, it can not select a time span to query, which is more laborious. Therefore only a small period of data could be used in the study, and this was not statistically significant (we will compare statistically with long-term data from SIDC later in Section 5). We searched for the daily sunspot groups numbers from April 1 to April 30, 2023, and calculated it in the same period by our algorithm. The results are shown in Table 1.

\begin{deluxetable*}{cchlDlc}
\tablenum{1}
\tablecaption{Compare the number of sunspot groups calculated by Our Algorithm with HFC’s \label{tab:messier}}
\tablewidth{0pt}
\tablehead{
\colhead{Date} & \colhead{Our algorithm}  & \multicolumn2c{HFC} & \colhead{Difference}
}
\startdata
20230401 & 5 && 5 & 0 \\
20230402 & 3 && 3 & 0 \\
20230403 & 3 && 2 & 1 \\
20230404 & 5 && 4 & 1 \\
20230405 & 4 && 3 & 1 \\
20230406 & 3 && 3 & 0 \\
20230407 & 4 && 3 & 1 \\
20230408 & 3 && 2 & 1 \\
20230409 & 4 && 3 & 1 \\
20230410 & 6 && 5 & 1 \\
20230411 & 5 && 5 & 0 \\
20230412 & 9 && 7 & 2 \\
20230413 & 9 && 7 & 2 \\
20230414 & 10 && 9 & 1 \\
20230415 & 14 && 10 & 4 \\
20230416 & 12 && 11 & 1 \\
20230417 & 10 && 7 & 3 \\
20230418 & 10 && 8 & 2 \\
20230419 & 10 && 8 & 2 \\
20230420 & 6 && 6 & 0 \\
20230421 & 9 && 6 & 3 \\
20230422 & 9 && 8 & 1 \\
20230423 & 6 && 7 & -1 \\
20230424 & 6 && 7 & -1 \\
20230425 & 9 && 6 & 3 \\
20230426 & 8 && 6 & 2 \\
20230427 & 5 && 8 & -3 \\
20230428 & 5 && 4 & 1 \\
20230429 & 4 && 4 & 0 \\
20230430 & 4 && 4 & 0 \\
\enddata
\end{deluxetable*}

As can be indicated in  Table 1, our results generally follow the same trend as those of HFC. However, the numbers of sunspot groups calculated by our algorithm were mostly larger than that of HFC. This was because our algorithm could detect weak sunspot groups that have just appeared or were about to disappear. The sunspot groups in the left image in Figure 3(a) were detected by our algorithm. The sunspot groups in the right image were marked by HFC. By comparison, the group numbered 3 was located at the edge of the solar disk, and it was found by our algorithm but ignored by HFC. In a few case, our algorithm mistook a large discrete sunspot group for two small groups. For example in Figure 3(b), the sunspot group numbered 2 and 3 should actually belong to a group. It was because our algorithm divided sunspots into a group according to the distance between sunspots. If the distance between sunspots in a large group was bigger than a fixed value, it would be divided into two groups.

\begin{figure*}
\gridline{\fig{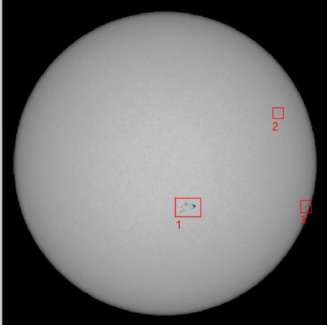}{0.35\textwidth}{(a)}
          \fig{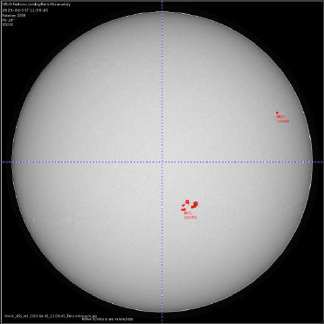}{0.35\textwidth}{(b)}
          }
\gridline{\fig{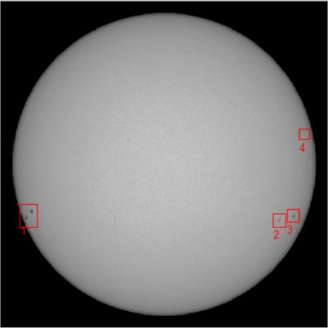}{0.35\textwidth}{(c)}
          \fig{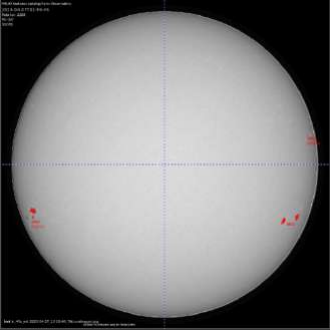}{0.35\textwidth}{(d)}
          }

\caption{Compare the sunspot grouping difference between our algorithm with HFC.
 On the left is the number of sunspot groups calculated by our algorithm on 2023-4-3. On the right is the number of sunspot groups marked by HFC on 2023-4-7.
\label{fig:figthree}}
\end{figure*}

\subsection{Accuracy and Efficiency} \label{subsec:results2}
To test the accuracy of our model, we set up a benchmark for comparison using the manual labelling results. As shown in Table 2, we manually annotated the data from April 1, 2023 to April 30, 2023 and then calculated four parameters: True Positive (TP), False Negative (FN), False Positive (FP), and True Negative (TN). They are the concepts from the Confusion matrix(\cite{visa2011confusion}). TP means that the sunspot groups on the solar disk are correctly identified and predicted to be positive. FN indicates that the originally existing sunspot group on the solar disk is not recognized, and is incorrectly predicted as the background area of the solar disk. FP indicates that the solar background or noisy regions are mistakenly identified as a sunspot group. TN represents the correct recognition of the solar background or noise regions as background, which is meaningless here.

\begin{deluxetable*}{cchlDlcccc}
\tablenum{2}
\tablecaption{Sunspot groups number by our model and by manual means \label{tab:tab2}}
\tablewidth{0pt}
\tablehead{\colhead{Date} & \colhead{Our Model}   & \multicolumn2c{Manual Means}  & \multicolumn2c{TP} & \colhead{FN} & \colhead{FP} & \colhead{TN}}
\startdata
20230401 & 5 && 5 && 5 & 0 & 0 & 0 \\
20230402 & 3 && 3 && 3 & 0 & 0 & 0 \\
20230403 & 3 && 3 && 3 & 0 & 0 & 0 \\
20230404 & 5 && 5 && 5 & 0 & 0 & 0 \\
20230405 & 4 && 4 && 4 & 0 & 0 & 0 \\
20230406 & 3 && 3 && 3 & 0 & 0 & 0 \\
20230407 & 4 && 3 && 2 & 1 & 2 & 0 \\
20230408 & 3 && 3 && 3 & 0 & 0 & 0 \\
20230409 & 4 && 4 && 4 & 0 & 0 & 0 \\
20230410 & 6 && 6 && 6 & 0 & 0 & 0 \\
20230411 & 5 && 5 && 5 & 0 & 0 & 0 \\
20230412 & 9 && 7 && 7 & 0 & 2 & 0 \\
20230413 & 9 && 9 && 9 & 0 & 0 & 0 \\
20230414 & 10 && 9 && 9 & 0 & 1 & 0 \\
20230415 & 14 && 11 && 11 & 0 & 3 & 0 \\
20230416 & 12 && 10 && 10 & 0 & 2 & 0 \\
20230417 & 10 && 8 && 8 & 0 & 2 & 0 \\
20230418 & 10 && 8 && 8 & 0 & 2 & 0 \\
20230419 & 10 && 8 && 8 & 0 & 2 & 0 \\
20230420 & 6 && 5 && 5 & 0 & 1 & 0 \\
20230421 & 9 && 7 && 7 & 0 & 2 & 0 \\
20230422 & 9 && 7 && 7 & 0 & 2 & 0 \\
20230423 & 6 && 5 && 5 & 0 & 1 & 0 \\
20230424 & 6 && 6 && 0 & 0 & 0 & 0 \\
20230425 & 9 && 7 && 7 & 0 & 2 & 0 \\
20230426 & 8 && 7 && 7 & 0 & 1 & 0 \\
20230427 & 5 && 5 && 0 & 0 & 0 & 0 \\
20230428 & 5 && 5 && 0 & 0 & 0 & 0 \\
20230429 & 4 && 4 && 0 & 0 & 0 & 0 \\
20230430 & 4 && 4 && 0 & 0 & 0 & 0 \\
Total    &   &&   && 151&1 & 25 & 0 \\
\enddata
\end{deluxetable*}

To evaluate the model, we used machine learning classification algorithm metrics such as Accuracy, Precision, Recall, and F1 Score, as shown in formulas (1) - (4)\citep{goutte2005probabilistic, yacouby2020probabilistic}. Accuracy refers to the "number of correctly predicted samples/total number of samples". Precision refers to how many of the samples we predict to be true are indeed true. Recall refers to how many samples that are actually true have been selected by the model. Precision and Recall are often mutually exclusive. The F1 score takes into account both Precision and Recall factors, achieving a compromise between them. The F1 score can be directly used to evaluate the performance of a model.

\begin{equation}
Accuracy = \frac{TP+TN}{TP+TN+FP+FN}
\end{equation}

\begin{equation}
Precision = \frac{TP}{TP+FP}
\end{equation}

\begin{equation}
Recall = \frac{TP}{TP+FN}
\end{equation}

\begin{equation}
F1 Score = 2 * \frac{Precision* Recall}{Precision+ Recall}
\end{equation}

After substituting the data in Table 2 into the equation, the results can be seen in Table 3. As indicated in the table, the performance of our model is significantly better than that of \cite{palladino2022sunspot}.\cite{palladino2022sunspot} proposed a method where a pipline of state-of-the-art sunspots groups detection and classification. Deep neural networks (DNNs) was adopted in their study to detect and classify sunspots groups in real time. They used the McIntosh system for sunspot groups classification and the generation process was carried out on the overall HMI data available on the timespan that went from 2010 to 2021 with a sampling time of 1 day. In order to classify sunspots groups, they paid more attention to the area sunspots groups in detection stage. Therefore, the accuracy metrics they choose is Intersection over Union (IoU) between predicted bounding box and ground truth bounding box. Intersection over union was then computed as the ratio of the overlapping area of the two bounding boxes over the union of their total area). Our choice of accuracy metrics was dependent on the number of sunspots groups, which was more reasonable as our purpose was to automatically calculate the relative SN.

As can be seen from the table, our algorithm has high recall rate of 99.3\%, indicating that almost all sunspot groups present on the solar disk can be accurately detected. The precision rate of 85.8\% is lower than the recall rate, which could explained by the fact that faint sunspot groups that were difficult to observe manually were now detectable by our algorithm.

\begin{deluxetable*}{cchlDlc}
\tablenum{3}
\tablecaption{Performance on Sunspot Group Detection. \label{tab:tab3}}
\tablewidth{0pt}
\tablehead{\colhead{Metric} & \multicolumn2c{\cite{palladino2022sunspot}}  & \colhead{Our Model}   }
\startdata
Precision & 65.82\% && 85.8\% \\
Recall    & 57.45\% && 99.3\% \\
F1 Score & 61.35\% && 92.1\%  \\
Accuracy & 44.22\% && 85.3\%  \\
\enddata
\end{deluxetable*}

In terms of computation time, our algorithm processed over 500 pieces of data on a single core regular desktop computer, in 5 minutes, with an average processing time of 0.6 seconds per data. In contrast, the algorithm proposed by \cite{palladino2022sunspot} initially took nearly 8 hours of computation time to generate nearly 3000 images on a regular desktop computer. After parallelizing on a 6-core CPU, the computational time was reduced to 1 hour. Our algorithm is clearly faster and has better real-time processing performance.

\section{Calculate the K value of HMI Relative Sunspot Number} \label{sec:K}
By writing a Python data collection program, we downloaded data between January 2022 to March 2023 and selected a daily sample of 507 images from \url{http://jsoc.stanford.edu/data/hmi/images/}. Then our algorithm was conducted to calculate the relative SN. On the other hand, with SIDC as calibration, we downloaded the SN in the same time span from \url{https://www.sidc.be/SILSO/home}. Then the correlation between them was calculated and shown in Figure 4. The horizontal axis represents the date (year-month), and the vertical axis denotes the Relative SN. The correlation coefficient between the two sets of data is 0.91. It is highly consistent, indicating that our algorithm is reliable.

\begin{figure}[ht!]
\plotone{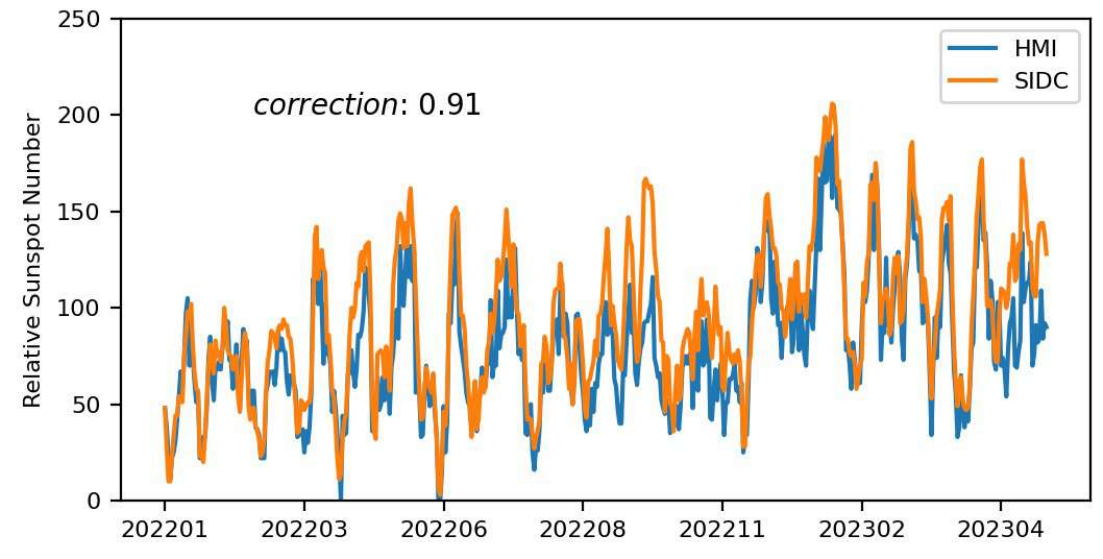}
\caption{Correlation between the relative SN of our algorithm and of SIDC  \label{fig:figfour}}
\end{figure}

As shown in Figure 5, we performed the least squares fit to both sets of data, which resulted in a fit of K=1.03. This is the scaling value of the HMI instrument.

\begin{figure}[ht!]
\plotone{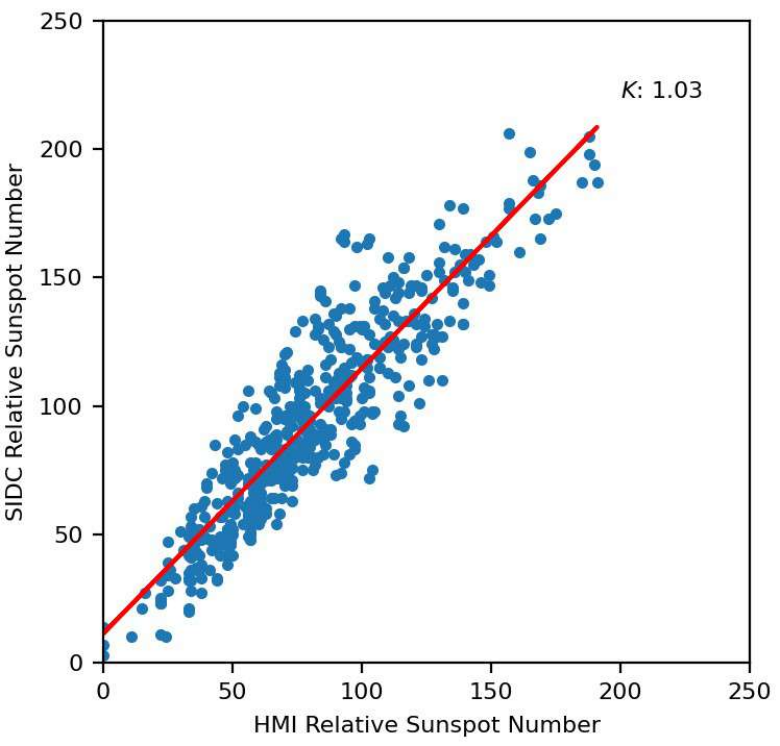}
\caption{Least squares fitting of the sunspot SN in HMI and SIDC to calculate the K value of HMI  \label{fig:figfive}}
\end{figure}

\section{Summary} \label{sec:Summary}
The long-term relative SN is statistically helpful in exploring the periodic patterns of the sun. In this paper, an automatic algorithm for calculating the relative SN is proposed. The experiments were conducted on HMI dataset. The conclusions of this article are as follows: (1) The performance of our algorithm: Accuracy=85.3\%, Precision=85.8\%, Recall=99.3\%, and F1 Score=92.1\%. (2) The average processing time of our algorithm is 0.6 seconds per data, which is remarkably fast and real-time. (3) The relative SN calculated by our algorithm is highly consistent with SIDC with a correlation of 0.91. This indicates our algorithm is reliable. (4) For the first time, we have calculated and confirmed the K value of SDO/HMI, which is K=1.03.

\begin{acknowledgments}

The authors would like to thank the anonymous reviewers for comments and suggestions that improved the quality of this manuscript. Images and data used in this paper are provided by SDO/HMI, SIDC, and HFC, for which we are very grateful. This work is supported by the Academic Research Projects of Beijing University (No. ZK20202204, ZK90202106, ZK90202105), the General projects of science and technology plan of Beijing Municipal Education Commission (KM20211141700) and the R\&D Program of Beijing Municipal Education Commission (KM201911417003, KM202211417006).

\end{acknowledgments}

%





\bibliography{myref}{}

\begin{thebibliography}{}
\expandafter\ifx\csname natexlab\endcsname\relax\def\natexlab#1{#1}\fi
\providecommand{\url}[1]{\href{#1}{#1}}
\providecommand{\dodoi}[1]{doi:~\href{http://doi.org/#1}{\nolinkurl{#1}}}
\providecommand{\doeprint}[1]{\href{http://ascl.net/#1}{\nolinkurl{http://ascl.net/#1}}}
\providecommand{\doarXiv}[1]{\href{https://arxiv.org/abs/#1}{\nolinkurl{https://arxiv.org/abs/#1}}}

\bibitem[{Bonnin {et~al.}(2013)Bonnin, Fuller, Reni{\'e}, Aboudarham, Cecconi, Bentley, \& Csillaghy}]{bonnin2013heliophysics}
Bonnin, X., Fuller, N., Reni{\'e}, C., {et~al.} 2013, Proceedings of the International Astronomical Union, 8, 512

\bibitem[{Carvalho {et~al.}(2020)Carvalho, Gomes, Barata, Louren{\c{c}}o, \& Peixinho}]{carvalho2020comparison}
Carvalho, S., Gomes, S., Barata, T., Louren{\c{c}}o, A., \& Peixinho, N. 2020, Astronomy and Computing, 32, 100385

\bibitem[{Clette {et~al.}(2007)Clette, Berghmans, Vanlommel, Van~der Linden, Koeckelenbergh, \& Wauters}]{clette2007wolf}
Clette, F., Berghmans, D., Vanlommel, P., {et~al.} 2007, Advances in Space Research, 40, 919

\bibitem[{Clette {et~al.}(2015)Clette, Cliver, Lef{\`e}vre, Svalgaard, \& Vaquero}]{clette2015revision}
Clette, F., Cliver, E., Lef{\`e}vre, L., Svalgaard, L., \& Vaquero, J. 2015, Space Weather, 13, 529

\bibitem[{Clette {et~al.}(2016)Clette, Lef{\`e}vre, Cagnotti, Cortesi, \& Bulling}]{clette2016revised}
Clette, F., Lef{\`e}vre, L., Cagnotti, M., Cortesi, S., \& Bulling, A. 2016, Solar Physics, 291, 2733

\bibitem[{Clette {et~al.}(2014)Clette, Svalgaard, Vaquero, \& Cliver}]{clette2014revisiting}
Clette, F., Svalgaard, L., Vaquero, J.~M., \& Cliver, E.~W. 2014, Space Science Reviews, 186, 35

\bibitem[{Curto {et~al.}(2008)Curto, Blanca, \& Mart{\'\i}nez}]{curto2008automatic}
Curto, J., Blanca, M., \& Mart{\'\i}nez, E. 2008, Solar Physics, 250, 411

\bibitem[{Dasgupta {et~al.}(2011)Dasgupta, Singh, \& Jewalikar}]{dasgupta2011sunspot}
Dasgupta, U., Singh, S., \& Jewalikar, V. 2011, in 2011 Third National Conference on Computer Vision, Pattern Recognition, Image Processing and Graphics, IEEE, 171--174

\bibitem[{Goutte \& Gaussier(2005)}]{goutte2005probabilistic}
Goutte, C., \& Gaussier, E. 2005, in European conference on information retrieval, Springer, 345--359

\bibitem[{Hanaoka(2022)}]{hanaoka2022automated}
Hanaoka, Y. 2022, Solar Physics, 297, 158

\bibitem[{Hathaway \& Wilson(2004)}]{hathaway2004sunspot}
Hathaway, D.~H., \& Wilson, R.~M. 2004, Solar Physics, 224, 5

\bibitem[{Heijmans(1995)}]{heijmans1995mathematical}
Heijmans, H. 1995, in Proceedings of Summer School on Morphological Image and Signal Processing, Citeseer, 228--231

\bibitem[{Hossfield(2002)}]{hossfield2002history}
Hossfield, C.~H. 2002, Journal of the American Association of Variable Star Observers, Vol. 31, No. 1, p. 48-53, 31, 48

\bibitem[{Lukianova \& Mursula(2011)}]{lukianova2011changed}
Lukianova, R., \& Mursula, K. 2011, Journal of Atmospheric and Solar-Terrestrial Physics, 73, 235

\bibitem[{Palladino {et~al.}(2022)Palladino, Ntagiou, Klug, Palacios, \& Keil}]{palladino2022sunspot}
Palladino, L., Ntagiou, E., Klug, J., Palacios, J., \& Keil, R. 2022, in 2022 IEEE Aerospace Conference (AERO), IEEE, 1--10

\bibitem[{Serra(1982)}]{serra1982image}
Serra, J. 1982, (No Title)

\bibitem[{Solanki {et~al.}(2004)Solanki, Usoskin, Kromer, Sch{\"u}ssler, \& Beer}]{solanki2004unusual}
Solanki, S.~K., Usoskin, I.~G., Kromer, B., Sch{\"u}ssler, M., \& Beer, J. 2004, Nature, 431, 1084

\bibitem[{Vanlommel {et~al.}(2004)Vanlommel, Cugnon, Linden, Berghmans, \& Clette}]{vanlommel2004sidc}
Vanlommel, P., Cugnon, P., Linden, R. V.~D., Berghmans, D., \& Clette, F. 2004, Solar Physics, 224, 113

\bibitem[{Visa {et~al.}(2011)Visa, Ramsay, Ralescu, \& Van Der~Knaap}]{visa2011confusion}
Visa, S., Ramsay, B., Ralescu, A.~L., \& Van Der~Knaap, E. 2011, Maics, 710, 120

\bibitem[{Wang(2004)}]{wang2004sun}
Wang, Y.-M. 2004, Solar Physics, 224, 21

\bibitem[{Yacouby \& Axman(2020)}]{yacouby2020probabilistic}
Yacouby, R., \& Axman, D. 2020, in Proceedings of the first workshop on evaluation and comparison of NLP systems, 79--91

\bibitem[{Yan {et~al.}(2011)Yan, Deng, Qu, \& Xu}]{yan2011phase}
Yan, X., Deng, L., Qu, Z., \& Xu, C. 2011, Astrophysics and Space Science, 333, 11

\bibitem[{Zhao {et~al.}(2016)Zhao, Lin, Deng, \& Yang}]{zhao2016automatic}
Zhao, C., Lin, G., Deng, Y., \& Yang, X. 2016, Publications of the Astronomical Society of Australia, 33, e018

\end{thebibliography}
\bibliographystyle{aasjournal}



\end{document}